\documentclass{optica-article}

\journal{opticajournal} 

\articletype{Research Article}
\usepackage{subcaption}
\usepackage{dblfloatfix}
\usepackage{lineno}
\usepackage{soul}

\begin{document}

\title{Experimental Evaluation of Passive Polarization Compensation Techniques for Fiber-Distributed Polarization-Entangled Photons}

\author{Gayatri Thik,\authormark{1} Amit Loyal,\authormark{1} Srinivasan K,\authormark{1,*} and Raghavan G\authormark{1}}

\address{\authormark{1}School of Quantum Technology, Defence Institute of Advanced Technology, Pune, India}

\email{\authormark{*}srinik@diat.ac.in} 


\begin{abstract*}
Entanglement distribution through optical fibers is essential for quantum communication networks; however, fiber transmission can alter photon polarization and modify the observed correlations of polarization-entangled states, necessitating polarization compensation to recover the desired entangled state in the measurement basis. Here, we experimentally evaluate two passive polarization compensation techniques: a free-space Quarter-Half-Quarter (QHQ) waveplate configuration and an in-fiber three-paddle Fiber Polarization Controller (FPC). The polarization transformation along each downconverted-photon path is independently compensated through a systematic path-by-path optimization procedure. Using both techniques, we recover high-quality polarization correlations and entanglement, with visibilities exceeding $93\%$ in three mutually unbiased bases and fidelities above $94.5\%$. The results demonstrate comparable restoration using free-space waveplate-based and fiber-based control, establishing a systematic framework for laboratory and short-reach quantum communication links.

\end{abstract*}

\section{Introduction}

Quantum entanglement is a key resource for quantum information processing and quantum communication, including quantum key distribution (QKD), quantum teleportation, entanglement swapping, and distributed quantum networking \cite{Gisin2002,Flamini2019}. Among photonic degrees of freedom, polarization-entangled photon pairs generated through spontaneous parametric down-conversion (SPDC) are widely used because of their high source brightness, straightforward manipulation using standard linear optics, and well-established projective measurement schemes \cite{Flamini2019,Lim2008}.

In practical quantum communication systems, polarization-entangled photons can be distributed through either free-space channels or optical fibers, depending on the application. Free-space links are mainly affected by atmospheric turbulence, beam divergence, and pointing errors, whereas single-mode optical fibers (SMFs) provide a compact and low-loss medium for the distribution of photonic quantum states \cite{Gisin2002,Lim2008}. During fiber transmission, photons experience both attenuation and polarization transformation. While attenuation reduces the number of transmitted photons, polarization transformation modifies the polarization state and can alter the polarization correlations between the distributed photon pairs \cite{Lim2008}.

The polarization transformation introduced by an optical fiber arises from uncontrolled birefringence associated with intrinsic core non-circularities, mechanical stress, bending, twisting, and ambient temperature variations \cite{Imai1988,Okoshi1985}. These effects introduce polarization rotations and relative phase shifts between orthogonal polarization modes, which can vary with the fiber conditions \cite{Imai1988,Ulrich1979}. Consequently, fiber-induced birefringence can alter the polarization correlations of the transmitted entangled photons, resulting in reduced or modified correlation visibility and changes in the Bell--CHSH parameter \cite{Lim2008,Stromberg2024}. Therefore, preserving or restoring the desired polarization state after fiber transmission is essential for reliable fiber-based quantum communication.

To recover the desired polarization state after fiber transmission, polarization compensation techniques can be broadly classified into active and passive approaches. Active compensation continuously monitors the polarization state and compensates for polarization changes using electronically controlled polarization elements and feedback systems \cite{Heismann1994,Chen2007,Xavier2009,Wu2022,Tan2024}. Such approaches are particularly suitable for dynamic communication channels but require additional polarization monitoring, control electronics, and feedback mechanisms, increasing the overall system complexity. In contrast, passive compensation employs polarization optics, such as waveplates or fiber polarization controllers, to compensate for the polarization transformation introduced by the fiber \cite{Lefevre1980,Simon1990,Stromberg2024}. These approaches are particularly attractive for laboratory experiments and short-reach fiber links because they can be implemented using standard optical components without continuous electronic feedback.

Systematic procedures for fiber-polarization compensation have previously been investigated for both classical and quantum communication systems. Experimental polarization compensation has been demonstrated for entanglement-based quantum key distribution \cite{Shi2021}, while other approaches have addressed random polarization drift and the recovery of polarization-entangled states \cite{Ramos2022,Zhou2025}. Analytical approaches for polarization-drift compensation in all-fiber QKD systems have also been demonstrated \cite{Mayboroda2024}. In particular, Strömberg \textit{et al.} presented a prescriptive two-basis method for compensating polarization transformations introduced by optical fibers using a fiber polarization controller together with tunable birefringence introduced by wave plates \cite{Stromberg2024}. Their method provides a systematic procedure for restoring individual polarization states after fiber transmission by optimizing the transformation using two non-orthogonal polarization states. In contrast to their combined use of a fiber polarization controller and wave plates, the present work independently investigates two passive compensation approaches, namely a QHQ waveplate configuration and a three-paddle fiber polarization controller. The general principle of polarization-state optimization is applied independently to each photon path, and the effectiveness of the compensation is subsequently evaluated through two-photon polarization correlations and quantum-state characterization.

In this work, we experimentally evaluate two passive polarization compensation techniques for fiber-distributed polarization-entangled photons: a free-space QHQ waveplate configuration and a three-paddle fiber polarization controller. The contribution of this work is not the introduction of a new polarization-control element, but the experimental demonstration and quantitative evaluation of a systematic path-by-path compensation procedure for fiber-distributed polarization-entangled photons. The polarization transformation introduced along each photon path is independently compensated by applying the same optimization principle to both the QHQ and FPC configurations. The effectiveness of the two implementations is evaluated through polarization-correlation measurements, Bell--CHSH measurements, and quantum-state tomography. The results demonstrate comparable recovery of high-quality polarization correlations and polarization-entangled states after fiber transmission, providing a practical and reproducible framework for path-by-path passive polarization compensation in laboratory and short-reach fiber-based quantum communication links.

\section{Polarization Evolution and Compensation in Optical Fiber}

\subsection{Polarization Evolution During Fiber Transmission}

The polarization state of a photon evolves during propagation through an optical fiber because of birefringence arising from fiber imperfections, mechanical stress, bending, twisting, and temperature variations \cite{Imai1988,Ulrich1979,Noda1986}. This birefringence introduces relative phase shifts and polarization rotations between the orthogonal polarization components, resulting in an unknown polarization transformation that can vary with the fiber condition and surrounding environment \cite{Imai1988,Noda1986,Lim2006}. For polarization-entangled photons, this transformation modifies the polarization state of each photon independently and can alter the observed polarization correlations in a fixed measurement basis \cite{Shtaif2011,Lim2008}.

The polarization transformation introduced by the optical fiber can be represented as a unitary operation acting on the input polarization state \cite{Simon1990,Braband2025},

\begin{equation}
|\psi_{\mathrm{out}}\rangle =
U_{\mathrm{fiber}}|\psi_{\mathrm{in}}\rangle,
\end{equation}

where $|\psi_{\mathrm{in}}\rangle$ and $|\psi_{\mathrm{out}}\rangle$ denote the input and output polarization states, respectively, and $U_{\mathrm{fiber}}\in SU(2)$ represents the unknown unitary transformation introduced by the optical fiber. For a two-photon entangled input state $|\psi_{\mathrm{in}}\rangle$, the transformation can be written as

\begin{equation}
|\psi_{\mathrm{out}}\rangle =
\left(U_{\mathrm{fiber(s)}}\otimes
U_{\mathrm{fiber(i)}}\right)|\psi_{\mathrm{in}}\rangle,
\end{equation}

where $U_{\mathrm{fiber(s)}}$ and $U_{\mathrm{fiber(i)}}$ represent the polarization transformations introduced along the signal and idler paths, respectively.

\subsection{Principle of Polarization Compensation}

The objective of polarization compensation is to recover the desired polarization state by compensating for the transformations introduced by the optical fibers. Since the signal and idler photons propagate through separate fiber paths, each photon experiences a local polarization transformation. Ideally, independent compensation transformations are applied to the two paths such that

\begin{equation}
\left(
U_{\mathrm{comp(s)}} U_{\mathrm{fiber(s)}}
\right)
\otimes
\left(
U_{\mathrm{comp(i)}} U_{\mathrm{fiber(i)}}
\right)
=
I_{\mathrm{s}}\otimes I_{\mathrm{i}},
\end{equation}

where $U_{\mathrm{comp(s)}}$ and $U_{\mathrm{comp(i)}}$ represent the compensation transformations applied to the signal and idler paths, respectively, and $I_{\mathrm{s}}$ and $I_{\mathrm{i}}$ denote the corresponding identity operations. Ideally,

\begin{equation}
U_{\mathrm{comp(s)}} =
U_{\mathrm{fiber(s)}}^{-1},
\qquad
U_{\mathrm{comp(i)}} =
U_{\mathrm{fiber(i)}}^{-1}.
\end{equation}

In practice, the polarization transformations introduced by the fibers are generally unknown and can vary with the fiber condition \cite{Imai1988,Stromberg2024}. Therefore, the compensation elements are adjusted experimentally to approximate the corresponding inverse transformations and recover the desired polarization state. In this work, this compensation is implemented using passive polarization-control elements.
\subsection{Passive Polarization Compensation Using Waveplates and Fiber Polarization Controller}

In this work, two passive polarization compensation techniques are investigated: a QHQ waveplate configuration and a three-paddle FPC \cite{Kidoh1981,Lefevre1980}. The QHQ arrangement provides a three-element realization of arbitrary $SU(2)$ polarization transformations, while a three-paddle FPC implements an analogous sequence of controllable fiber retarders through stress-induced birefringence \cite{Simon1990,Lefevre1980}.

The QHQ waveplate combination compensates the polarization transformation by introducing controlled phase shifts between orthogonal polarization components. By appropriately adjusting the orientations of the three waveplates, the desired polarization transformation can be realized \cite{Simon1990,Reddy2016}.

Similarly, the three-paddle FPC introduces controlled birefringence through three loops of optical fiber wound around rotating paddles \cite{Shimizu1991,Lefevre1980}. The paddles provide a sequence of controllable fiber retarders that can be configured to implement a QHQ waveplate-like polarization transformation \cite{Simon1990}. Although both techniques provide controllable polarization transformations, they differ in their physical implementation: the QHQ configuration performs the compensation in free space, whereas the FPC performs it directly within the optical fiber.

The experimental procedure used to optimize these two compensation configurations is described in the following subsection.

\subsection{Experimental Procedure for Passive Polarization Compensation}
\label{sec:polarization_compensation}

The passive polarization-compensation procedure used in this work is based on independent, path-by-path optimization of the polarization transformations introduced by the individual fibers before characterization of the distributed entangled state. The procedure follows the general principle of preparing selected polarization states and adjusting the compensation elements to recover the corresponding desired output states \cite{Stromberg2024,Reddy2016}.

During fiber transmission, the signal and idler photons experience local polarization transformations along their respective paths. Consequently, the polarization transformation associated with each path can be independently optimized using single-photon polarization measurements before the compensated two-photon state is characterized.

The same optimization principle was applied to both compensation techniques. For the QHQ waveplate configuration, the waveplates were adjusted sequentially in the order $Q_2 \rightarrow \mathrm{HWP} \rightarrow Q_1$. The second quarter-wave plate ($Q_2$), placed closest to the analyzer, was adjusted first to reduce the ellipticity of the output polarization state. The half-wave plate (HWP) was then rotated to align the polarization with the required linear polarization basis. Finally, the first quarter-wave plate ($Q_1$), placed closest to the fiber output, was adjusted to compensate for the remaining phase difference between the orthogonal polarization components. Small iterative adjustments of the three waveplates were repeated until no further improvement in the polarization visibility was observed.

For the three-paddle FPC, the paddles were adjusted iteratively according to the same optimization principle until the maximum polarization visibility was obtained.

The polarization compensation procedure was implemented as follows:

\begin{enumerate}

    \item A linear polarizer was temporarily inserted in the selected photon path to prepare a known input polarization state. A second linear polarizer placed after the compensation stage was used as the polarization analyzer.

    \item The polarization transformation associated with each optical path was optimized independently using measurements in the $H$, $V$, $D$, and $A$ polarization states. For each prepared state, the transmitted intensity was measured for the corresponding analyzer setting and its orthogonal setting. The use of multiple non-orthogonal polarization states allows the optimization to account for both polarization rotations and relative phase changes introduced by the fiber.

    \item The compensation elements were adjusted iteratively to maximize the transmission in the desired analyzer setting while minimizing the transmission in the corresponding orthogonal setting. The optimization was continued until no further improvement in the extinction ratio between the desired and orthogonal analyzer outputs was observed. For the QHQ configuration, the adjustment followed the sequence $Q_2 \rightarrow \mathrm{HWP} \rightarrow Q_1$. The three-paddle FPC was adjusted iteratively according to the same optimization principle.

    \item The same procedure was then independently applied to the second photon path.

    \item After optimization of both paths, the temporary polarizers were removed, and the distributed polarization-entangled photon pairs were characterized using polarization-correlation measurements, Bell--CHSH measurements, and quantum-state tomography.

\end{enumerate}

\section{Experimental Setup}

\begin{figure*}[!t]
    \centering

    \begin{subfigure}{0.75\textwidth}
        \centering
        \includegraphics[width=\textwidth]{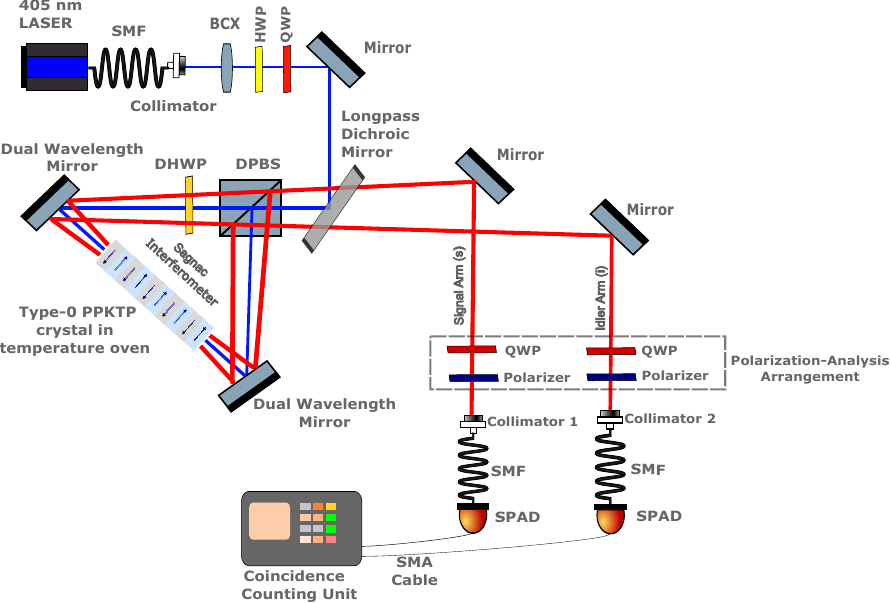}
        \caption{}
        \label{fig:main_source}
    \end{subfigure}

    \vspace{0.5em}

    \begin{subfigure}{1.0\textwidth}
        \centering
        \includegraphics[width=\textwidth]{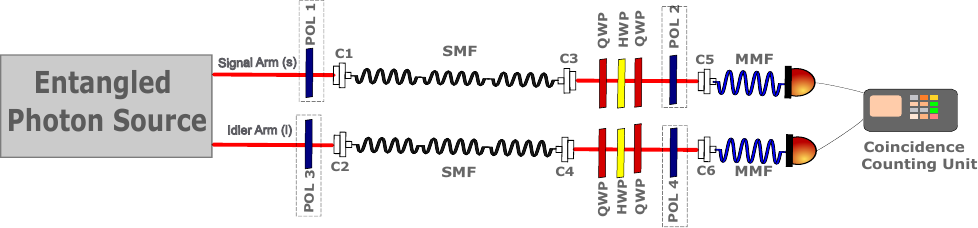}
        \caption{}
        \label{fig:qhq_correction}
    \end{subfigure}

    \vspace{0.5em}

    \begin{subfigure}{1.0\textwidth}
        \centering
        \includegraphics[width=\textwidth]{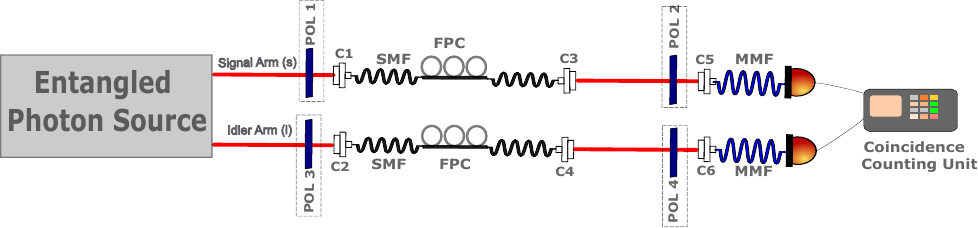}
        \caption{}
        \label{fig:fpc_correction}
    \end{subfigure}

    \caption{Experimental configurations for the generation, fiber distribution, passive polarization compensation, and characterization of polarization-entangled photon pairs. (a) Polarization-entangled photon source (EPS) and polarization-analysis arrangement used for reference characterization before fiber distribution. (b) QHQ-based polarization-compensation configuration, in which a free-space QWP--HWP--QWP sequence is introduced after the fiber output in each photon path. (c) FPC-based polarization-compensation configuration, in which the signal and idler distribution fibers are independently controlled using three-paddle fiber polarization controllers. In (b) and (c), the common EPS and photon-distribution arrangement are represented as a black-box source. The highlighted regions indicate the temporary polarizers (POL1--POL4) used during the path-by-path polarization optimization. These polarizers are removed after compensation, and the polarization-analysis elements are inserted for characterization of the compensated two-photon state.}

    \label{fig:experimental_setup}
\end{figure*}

The experimental setup used to evaluate the two passive polarization compensation techniques is shown in Fig.~\ref{fig:experimental_setup}. The setup consists of the polarization-entangled photon source (EPS), the fiber-based distribution and polarization-compensation stage, and the polarization-analysis and detection system. Figure~\ref{fig:experimental_setup}(a) shows the complete EPS and polarization-analysis arrangement used for reference source characterization, while Figs.~\ref{fig:experimental_setup}(b) and \ref{fig:experimental_setup}(c) show the QHQ- and FPC-based compensation configurations, respectively. For simplicity, the EPS is represented as a common source block in Figs.~\ref{fig:experimental_setup}(b) and \ref{fig:experimental_setup}(c). Except for the compensation elements, the optical arrangement was kept identical for both configurations, and the compensation was performed independently along the signal and idler paths.

\subsection{Polarization-Entangled Photon Source}

Polarization-entangled photon pairs were generated using a type-0 periodically poled potassium titanyl phosphate (PPKTP) crystal in a polarization Sagnac interferometer, as shown in Fig.~\ref{fig:experimental_setup}(a). The PPKTP crystal, with dimensions of $2\times1\times30$~mm and a poling period of $3.425~\mu$m, was maintained at $26.5~^{\circ}$C. A continuous-wave 405~nm laser was coupled through a single-mode fiber, collimated, and passed through a HWP and QWP to prepare the pump in a diagonal polarization state before being focused at the center of the crystal using a BCX lens.

A dual-wavelength polarizing beam splitter (DPBS) and a dual-wavelength half-wave plate (DHWP) directed the diagonally polarized pump into the two counter-propagating directions of the Sagnac interferometer. The SPDC photons generated in the two directions were recombined at the DPBS, producing the coherent superposition required for polarization entanglement. With the pump prepared in the diagonal polarization state, the source generated the target polarization-entangled state

\begin{equation}
|\Phi^+\rangle =
\frac{1}{\sqrt{2}}
\left(|HH\rangle+|VV\rangle\right).
\end{equation}

A longpass dichroic mirror separated the 405~nm pump from the downconverted photons around 810~nm, which were directed into the signal and idler paths. The downconverted photons were filtered using $810\pm10$~nm band-pass filters.

The generated state was first characterized directly at the EPS output to obtain a reference measurement before fiber distribution. The POL and QWP polarization-analysis elements in the two paths were used to perform the required polarization projections, and the corresponding coincidence measurements were recorded.

\subsection{Fiber Distribution, Polarization Compensation, and Characterization}

For fiber distribution, the signal and idler photons from the EPS were coupled into $2$~m single-mode fibers using collimators C1 and C2 [Figs.~\ref{fig:experimental_setup}(b) and \ref{fig:experimental_setup}(c)]. The fibers were kept fixed throughout the measurements. At the fiber outputs, C3 and C4 were used to collimate the photons into the respective free-space paths. The polarization state after fiber transmission was initially characterized using the POL and QWP polarization-analysis elements before applying polarization compensation.

For the QHQ configuration, a QWP--HWP--QWP combination was placed in the free-space section of each photon path after the fiber output. The fibers were kept fixed on the optical table throughout the measurements to maintain a stable polarization transformation. For the FPC configuration, the distribution fibers were placed inside the corresponding three-paddle FPCs and initially aligned with the reference fast axis of the controllers. Temporary polarizers were introduced before and after the compensation stage in each path to prepare and analyze the polarization states during the optimization procedure. The compensation elements in the signal and idler paths were optimized independently following the procedure described in Section~\ref{sec:polarization_compensation}.

After compensation, the temporary polarizers (POL1--POL4) were removed, and the polarization-analysis elements (POL and QWP) were arranged in the respective paths to characterize the distributed entangled state. The photons were subsequently collected using C5 and C6, coupled into multimode fibers (MMFs), and detected using silicon single-photon avalanche photodiodes (SPADs). The detector outputs were connected to the coincidence counting unit through SMA cables, and coincidence counts were recorded for the required polarization projections.

The distributed state was characterized before and after polarization compensation using the same polarization-analysis and detection arrangement. The measured coincidence counts were used to evaluate the polarization correlations and Bell--CHSH parameter, while two-qubit quantum state tomography was performed to determine the fidelity of the distributed state with respect to the target $\lvert\Phi^+\rangle$ state. The same source, fiber length, collection optics, and detection system were used for both compensation configurations, with the polarization-compensation element being the only experimental difference between them.

\section{Experimental Results and Discussion}

The performance of the path-by-path passive polarization-compensation procedure was evaluated under similar experimental conditions using the same polarization-entangled photon source and fiber paths. The generated $\left|\Phi^{+}\right\rangle$ state was first characterized at the source to establish a reference. The photons were then transmitted through the fibers, and the resulting two-photon state was characterized before compensation. The signal and idler paths were subsequently optimized independently using either a QHQ waveplate configuration or a three-paddle fiber polarization controller (FPC). The compensated two-photon state was finally evaluated through polarization-correlation measurements, Bell--CHSH measurements, and quantum state tomography (QST).

\subsection{Reference Characterization of the Entangled Photon Source}

\begin{figure}[!t]
\centering
\subfloat[Polarization correlations]{%
\includegraphics[width=0.48\columnwidth]{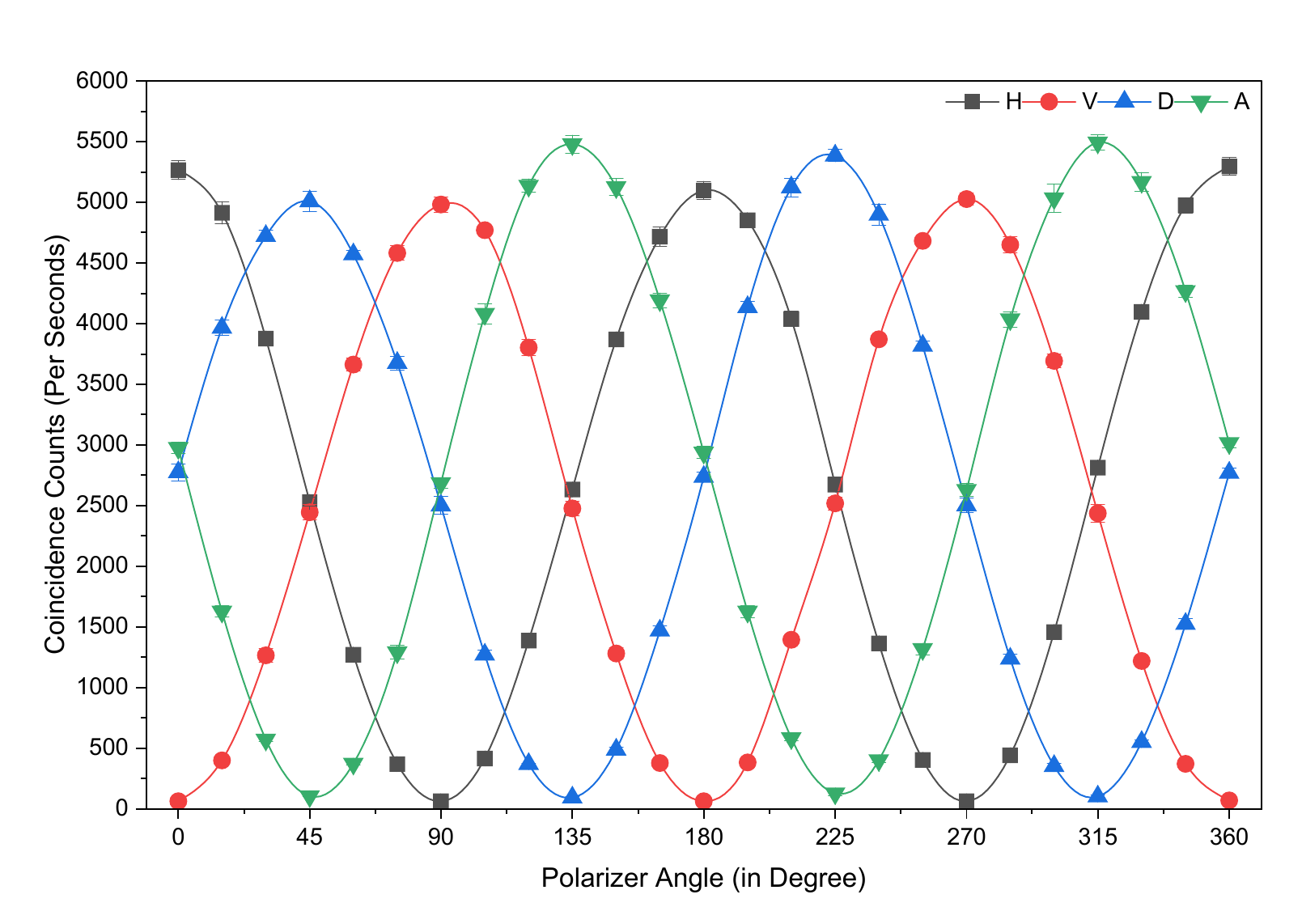}}
\hfill
\subfloat[Reconstructed density matrix]{%
\includegraphics[width=0.48\columnwidth]{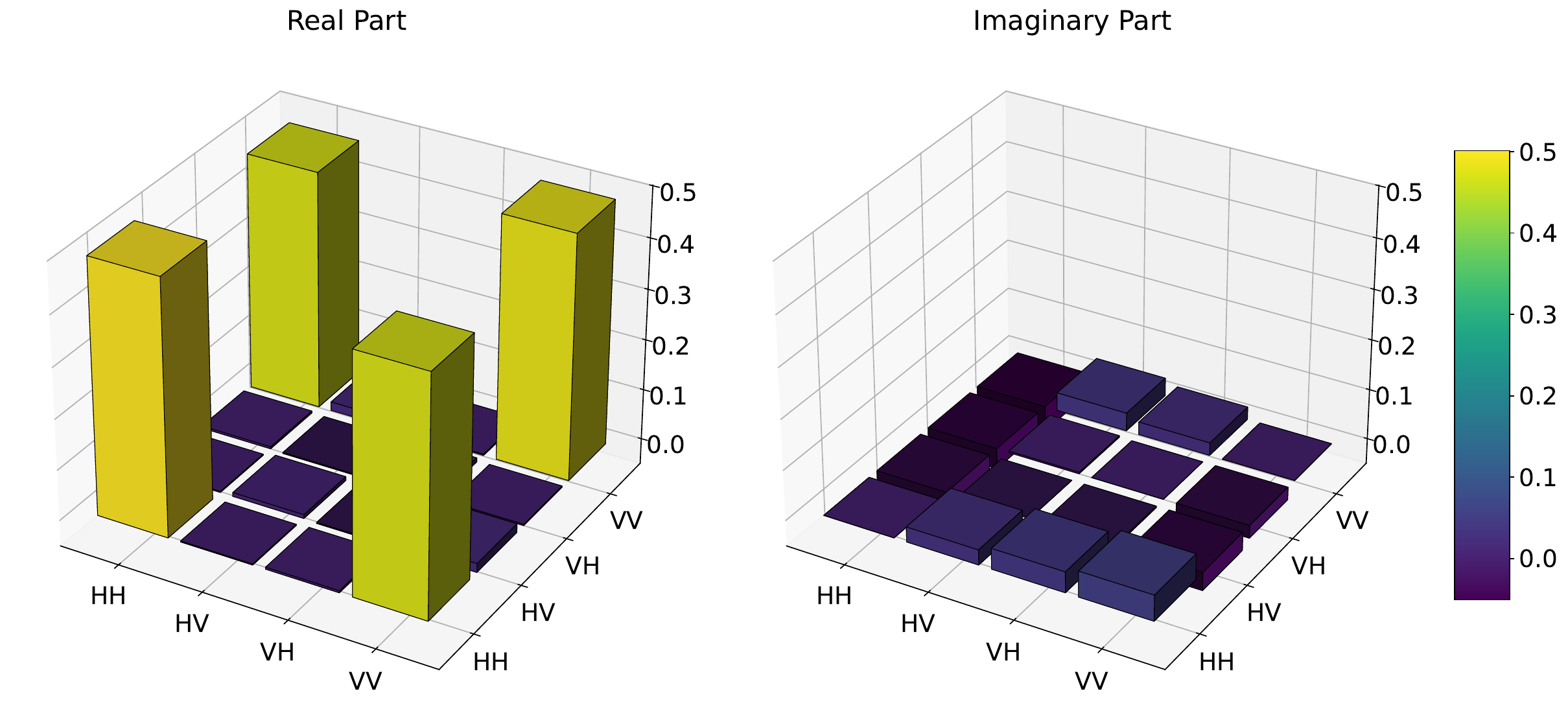}}

\caption{Reference characterization of the polarization-entangled photon source before fiber distribution. (a) Measured polarization correlations of the generated $\left|\Phi^{+}\right\rangle$ state in the $H/V$ and $D/A$ measurement bases. (b) Reconstructed two-qubit density matrix obtained from quantum-state tomography.}
\label{fig:reference_results}
\end{figure}

The polarization-entangled state generated by the EPS was first characterized before fiber distribution. The measured polarization correlations and reconstructed density matrix are shown in Fig.~\ref{fig:reference_results}. High polarization visibilities were obtained in the three measurement bases, as summarized in Table~\ref{tab:combined_results}. The measured CHSH parameter was $S=2.755\pm0.014$, demonstrating a clear violation of the classical bound. Quantum state tomography showed a fidelity of $96.12\%$ with respect to the target $\left|\Phi^{+}\right\rangle$ state, together with high concurrence and purity.

These measurements establish the quality of the generated entangled state and provide a reference for evaluating the polarization transformations introduced by fiber transmission and the subsequent recovery achieved through passive polarization compensation.

\begin{table*}[!t]
\centering
\caption{Characterization results of the reference polarization-entangled photon source and the distributed entangled photons before and after polarization compensation using the QHQ and FPC configurations.}
\label{tab:combined_results}

\begin{tabular}{lccccc}
\hline
\textbf{Parameter} &
\textbf{EPS Reference} &
\multicolumn{2}{c}{\textbf{QHQ}} &
\multicolumn{2}{c}{\textbf{FPC}} \\
& & \textbf{Before} & \textbf{After} & \textbf{Before} & \textbf{After} \\
\hline

Visibility, $H/V$
& $97.43\%$
& $17.36\%$
& $95.81\%$
& $16.42\%$
& $94.88\%$ \\

Visibility, $D/A$
& $97.38\%$
& $36.93\%$
& $95.85\%$
& $33.14\%$
& $95.49\%$ \\

Visibility, $R/L$
& $96.46\%$
& $14.63\%$
& $93.63\%$
& $40.01\%$
& $93.79\%$ \\

CHSH parameter, $|S|$
& $2.755 \pm 0.014$
& $1.093 \pm 0.020$
& $2.661 \pm 0.018$
& $0.249 \pm 0.019$
& $2.664 \pm 0.017$ \\

State fidelity
& $96.12\%$
& $9.18\%$
& $94.59\%$
& $34.27\%$
& $94.93\%$ \\

Concurrence
& $94.30\%$
& $89.23\%$
& $91.85\%$
& $94.12\%$
& $93.32\%$ \\

Purity
& $93.97\%$
& $88.50\%$
& $92.36\%$
& $94.09\%$
& $93.61\%$ \\

\hline
\end{tabular}
\end{table*}

\subsection{Polarization Compensation Using QHQ and FPC}

\begin{figure*}[!t]
\centering

\begin{subfigure}{0.48\textwidth}
    \centering
    \includegraphics[width=\linewidth]{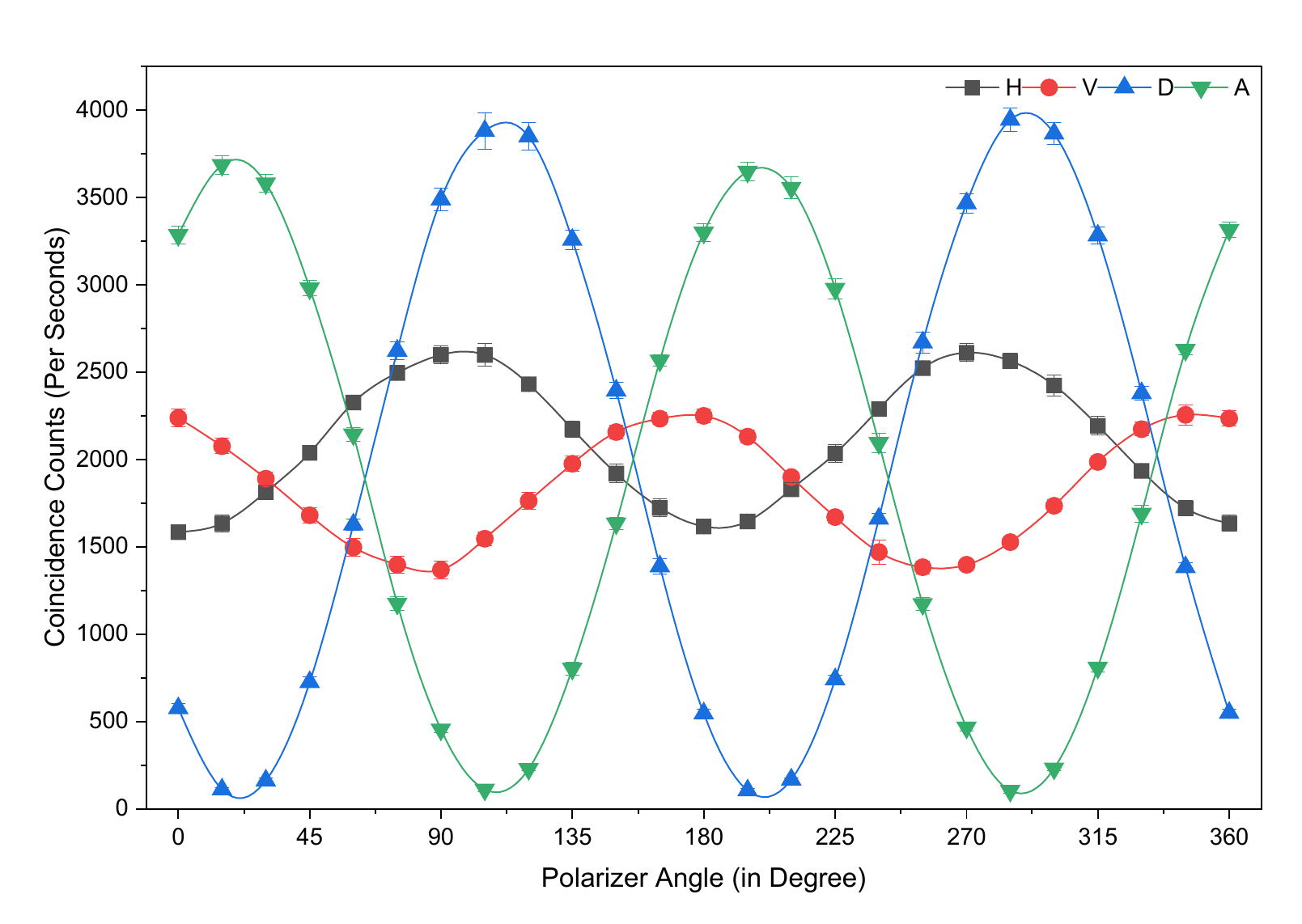}
    \caption{}
    \label{fig:qhq_before_corr}
\end{subfigure}
\hfill
\begin{subfigure}{0.48\textwidth}
    \centering
    \includegraphics[width=\linewidth]{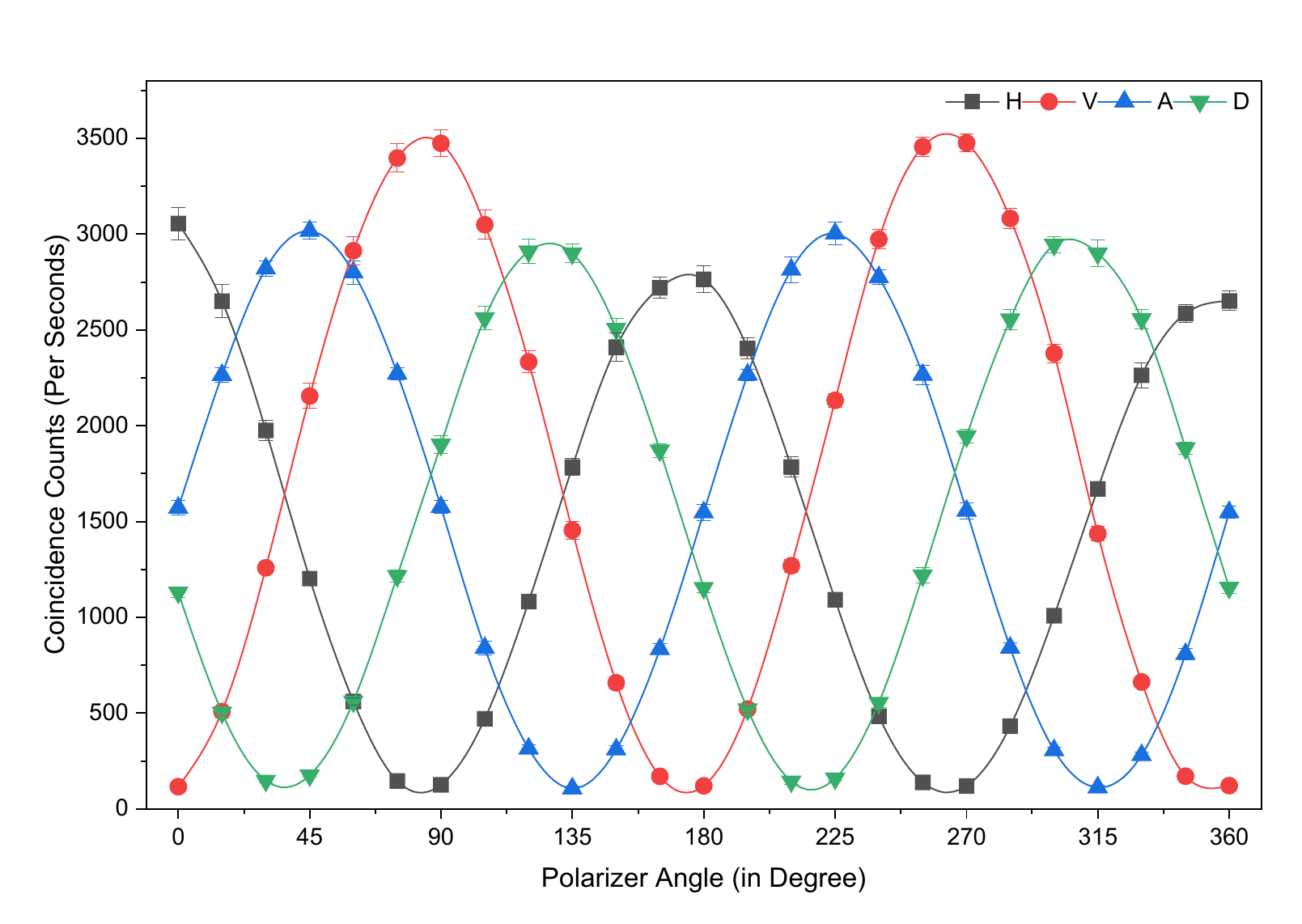}
    \caption{}
    \label{fig:qhq_after_corr}
\end{subfigure}

\vspace{0.4em}

\begin{subfigure}{0.48\textwidth}
    \centering
    \includegraphics[width=\linewidth]{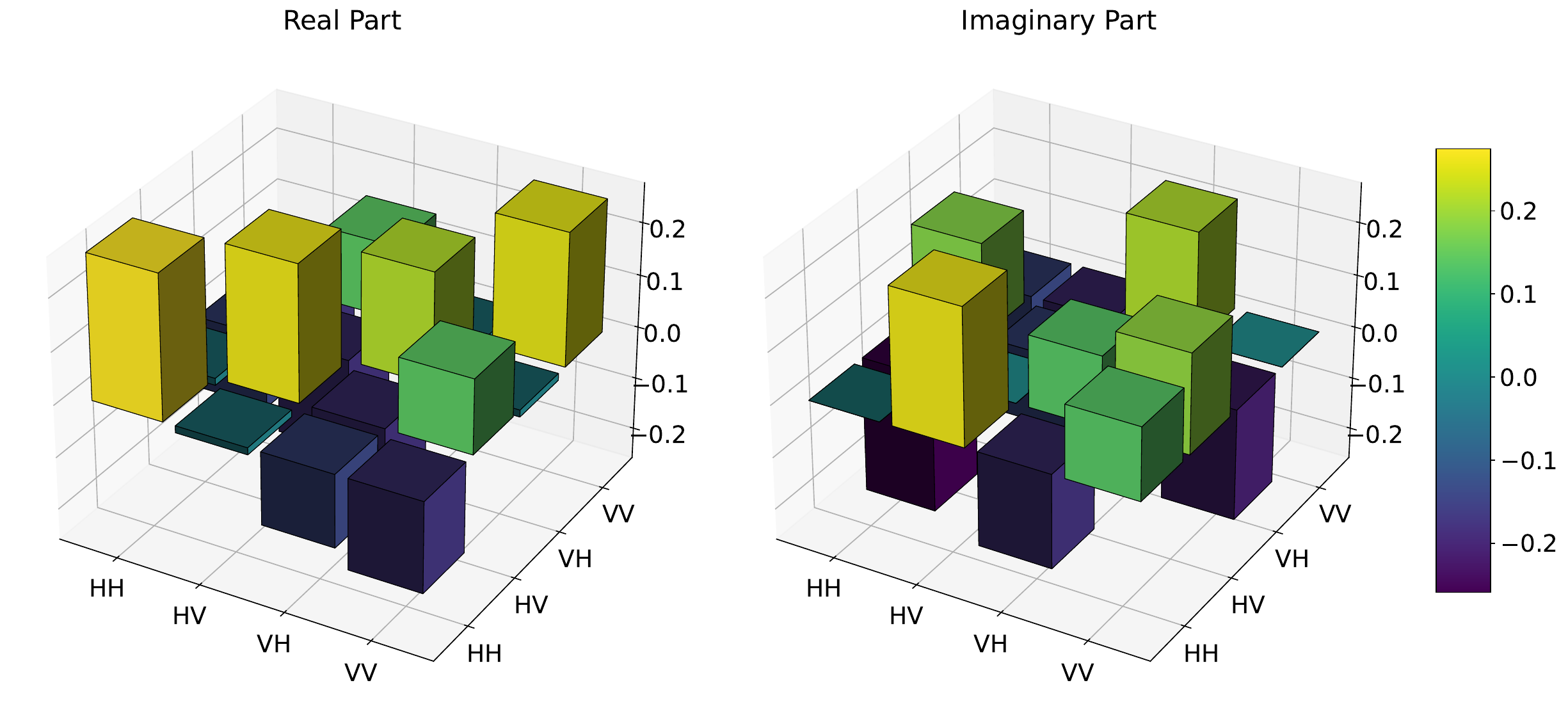}
    \caption{}
    \label{fig:qhq_before_dm}
\end{subfigure}
\hfill
\begin{subfigure}{0.48\textwidth}
    \centering
    \includegraphics[width=\linewidth]{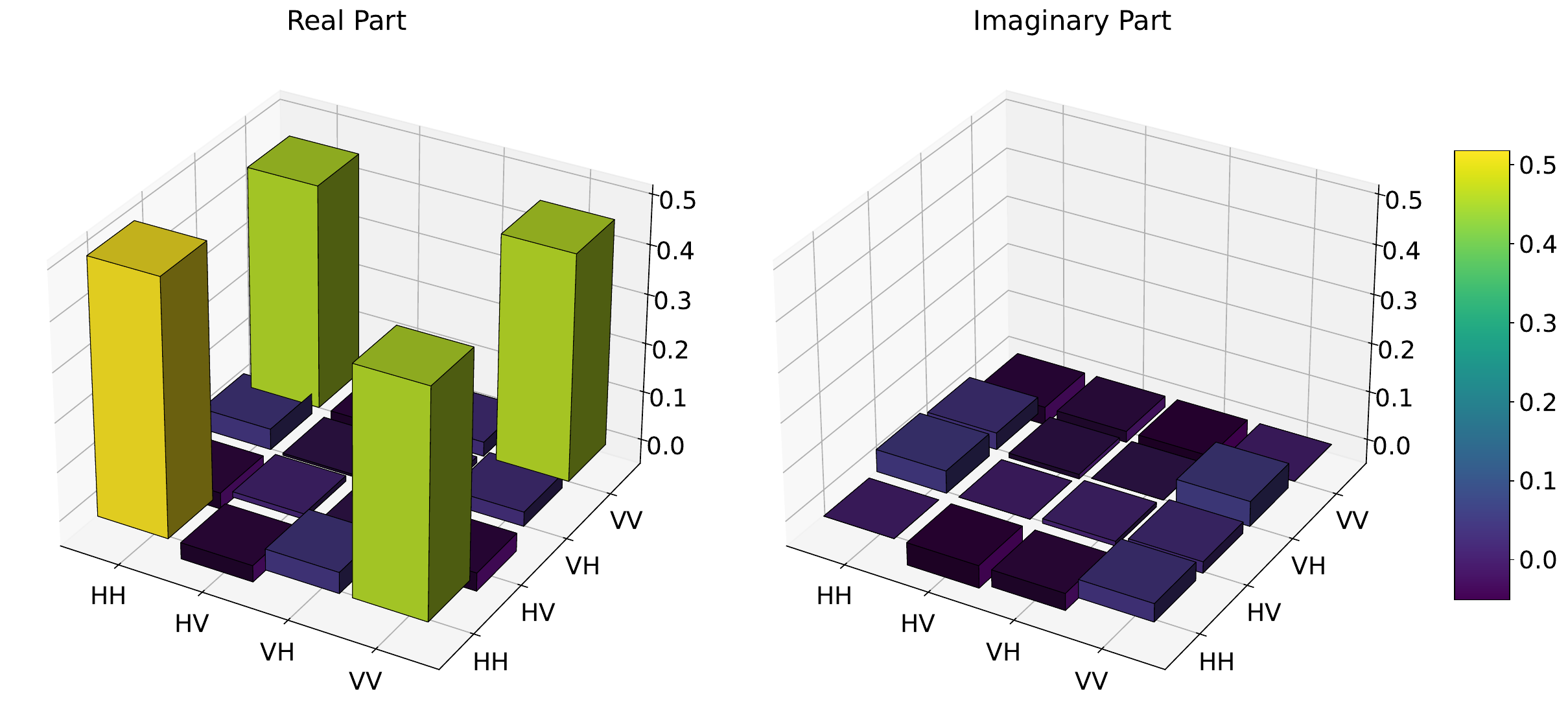}
    \caption{}
    \label{fig:qhq_after_dm}
\end{subfigure}

\caption{Polarization-correlation and quantum-state-tomography results before and after QHQ-based polarization compensation. (a) Polarization correlations before compensation. (b) Polarization correlations after path-by-path QHQ optimization. (c) Reconstructed density matrix before compensation. (d) Reconstructed density matrix after compensation, showing recovery of a state close to $\left|\Phi^{+}\right\rangle$.}
\label{fig:qhq_results}
\end{figure*}

\begin{figure*}[!t]
\centering

\begin{subfigure}{0.48\textwidth}
    \centering
    \includegraphics[width=\linewidth]{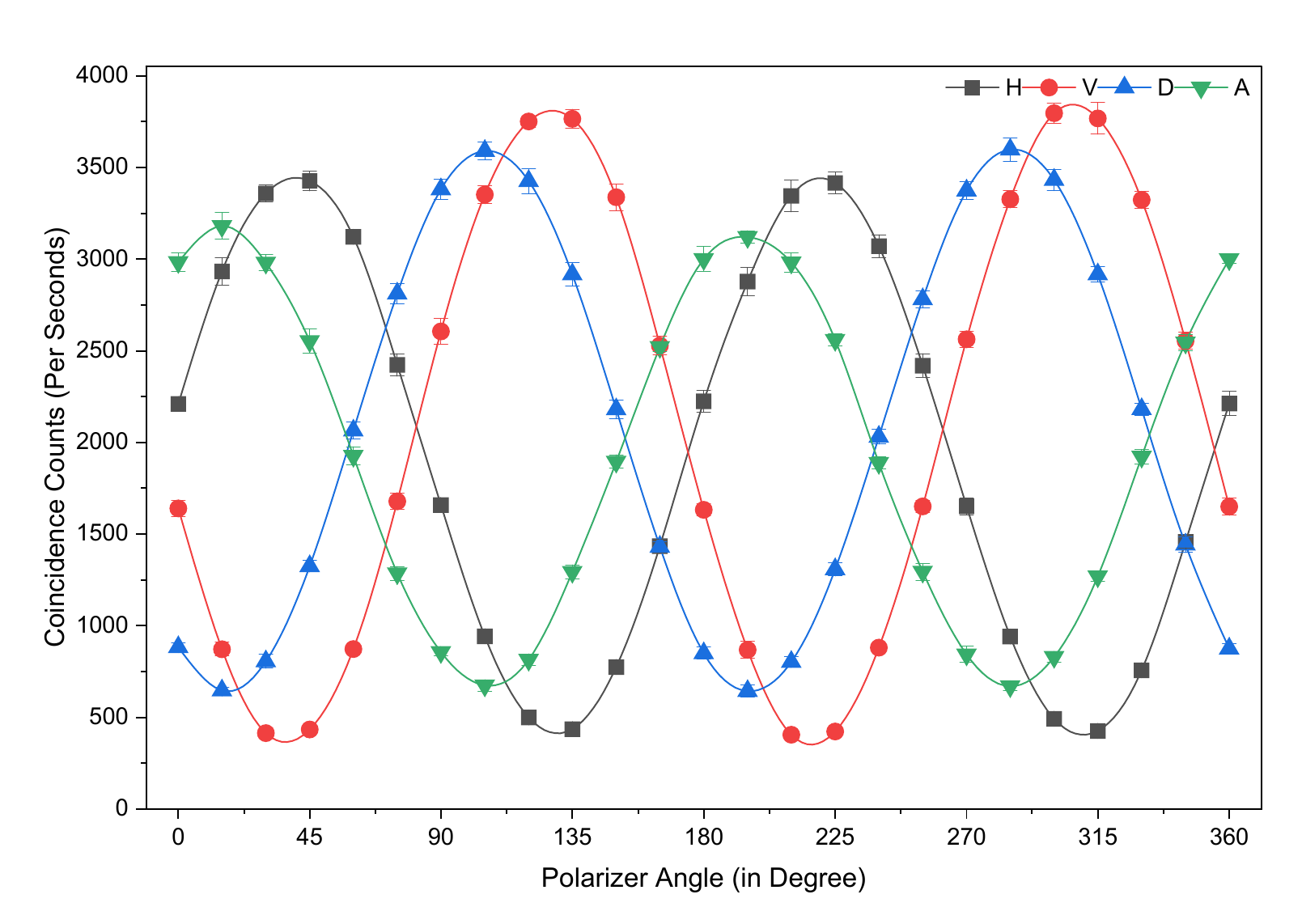}
    \caption{}
    \label{fig:fpc_before_corr}
\end{subfigure}
\hfill
\begin{subfigure}{0.48\textwidth}
    \centering
    \includegraphics[width=\linewidth]{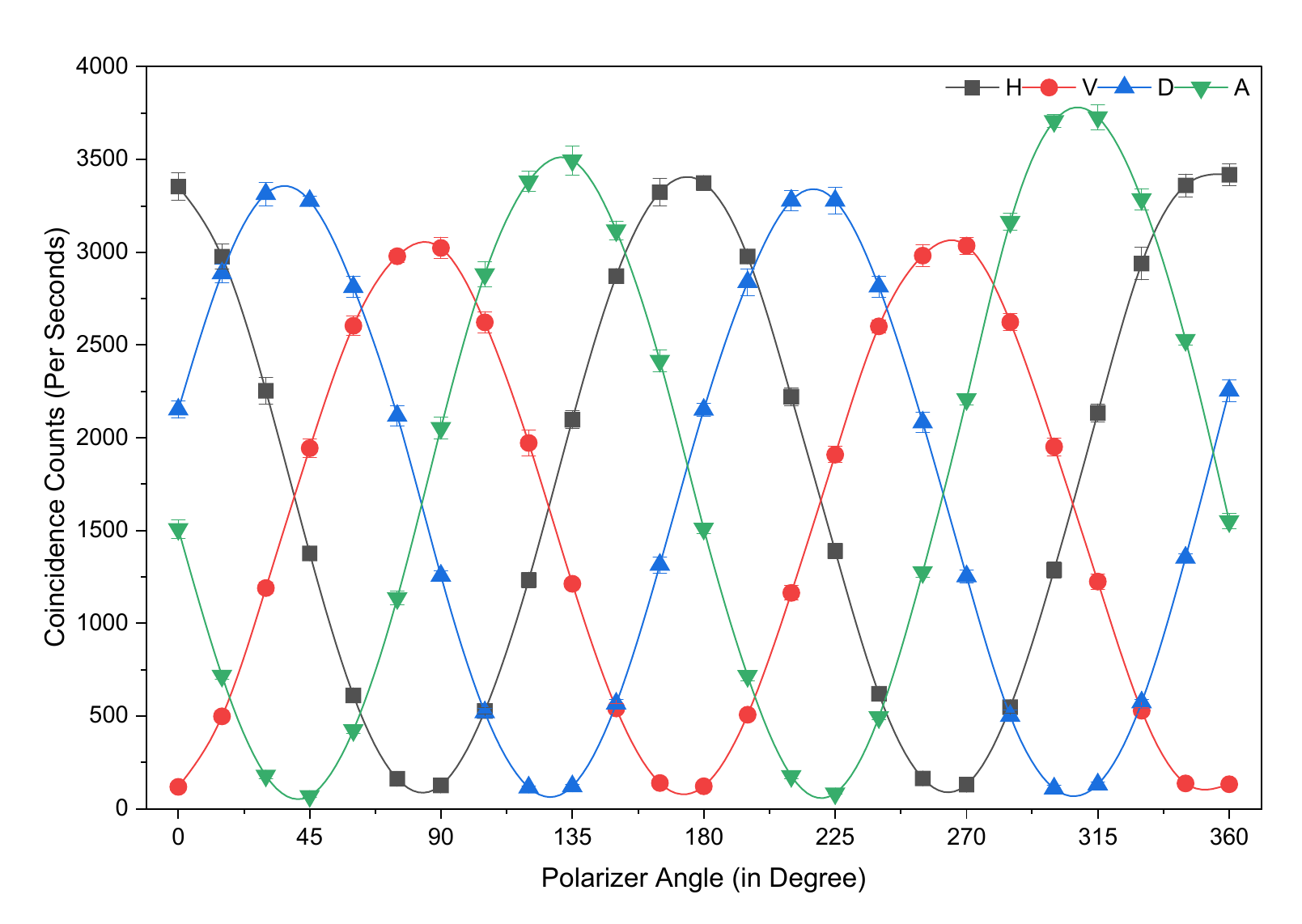}
    \caption{}
    \label{fig:fpc_after_corr}
\end{subfigure}

\vspace{0.4em}

\begin{subfigure}{0.48\textwidth}
    \centering
    \includegraphics[width=\linewidth]{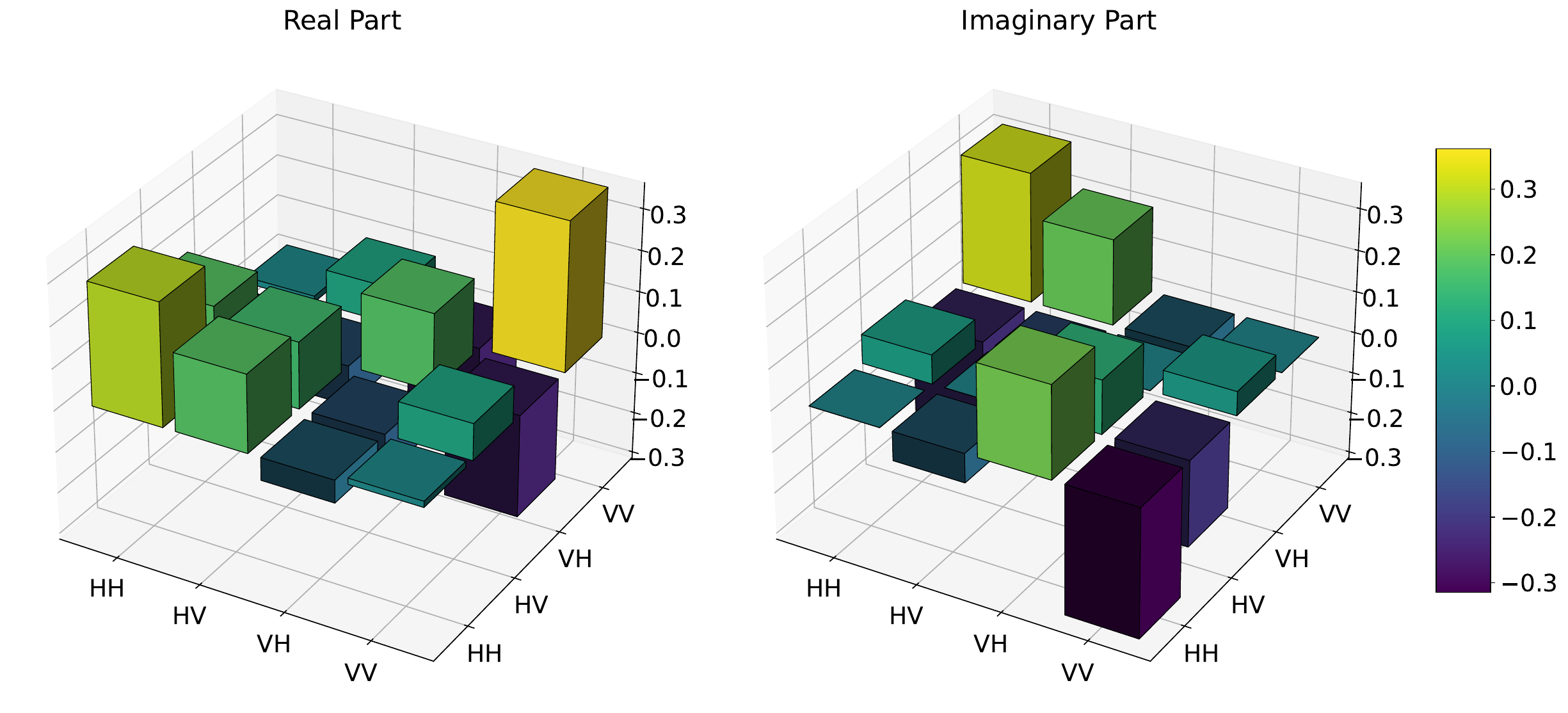}
    \caption{}
    \label{fig:fpc_before_dm}
\end{subfigure}
\hfill
\begin{subfigure}{0.48\textwidth}
    \centering
    \includegraphics[width=\linewidth]{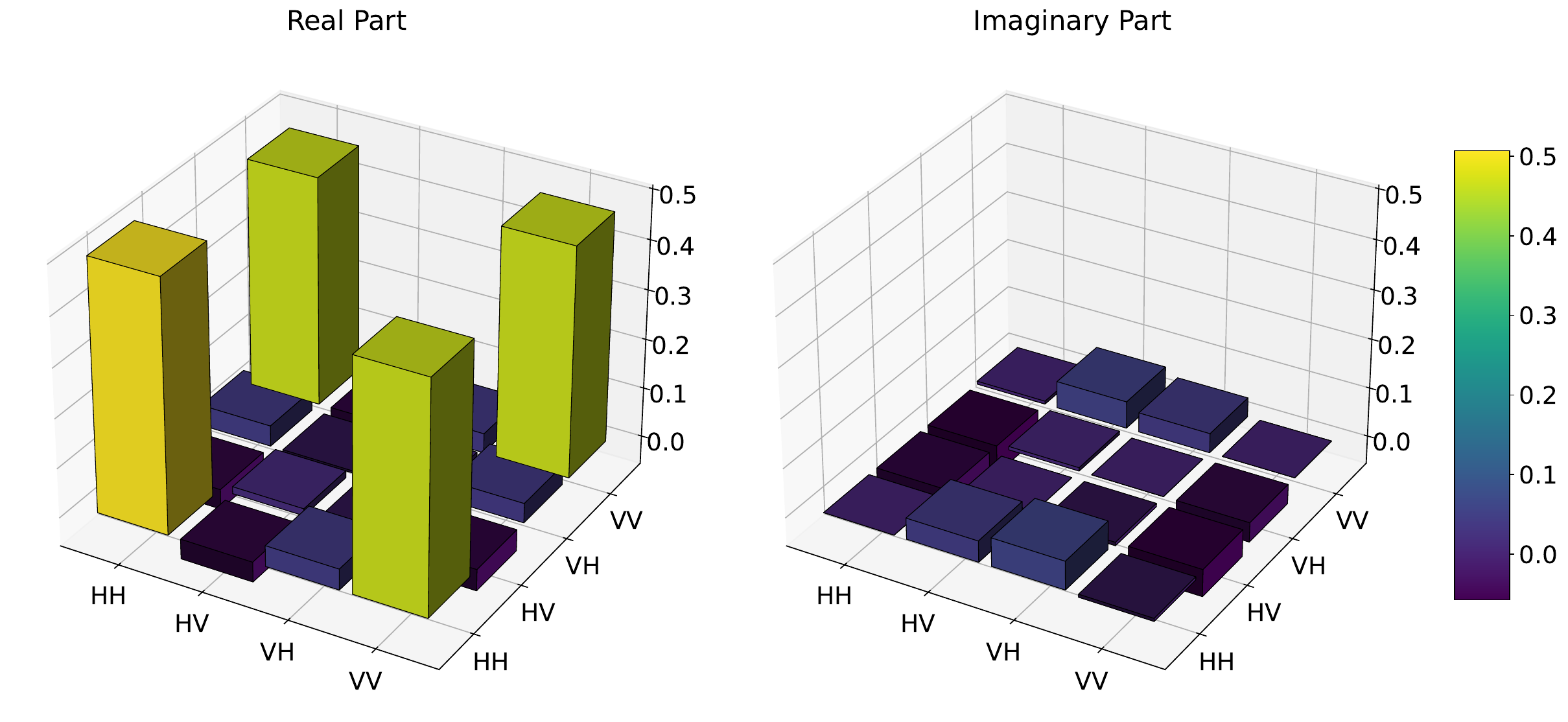}
    \caption{}
    \label{fig:fpc_after_dm}
\end{subfigure}

\caption{Polarization-correlation and quantum-state-tomography results before and after FPC-based polarization compensation. (a) Polarization correlations before compensation. (b) Polarization correlations after path-by-path FPC optimization. (c) Reconstructed density matrix before compensation. (d) Reconstructed density matrix after compensation, showing recovery of a state close to $\left|\Phi^{+}\right\rangle$.}
\label{fig:fpc_results}
\end{figure*}

The polarization correlations before and after compensation were evaluated using the visibility

\begin{equation}
V=\frac{C_{\mathrm{max}}-C_{\mathrm{min}}}
{C_{\mathrm{max}}+C_{\mathrm{min}}},
\end{equation}

where $C_{\mathrm{max}}$ and $C_{\mathrm{min}}$ denote the maximum and minimum coincidence counts obtained from the corresponding polarization projections, respectively. The visibility therefore quantifies the contrast between the maximum and minimum coincidence counts, with higher values indicating stronger polarization correlations.

The measurements labeled ``Before'' in Table~\ref{tab:combined_results} for the QHQ and FPC configurations were obtained separately before applying the respective polarization-compensation procedure. Although the same fiber paths and experimental arrangement were used, the measurements were performed independently. Therefore, the uncompensated polarization transformations did not necessarily produce identical visibility and correlation values in the two configurations.

Before polarization compensation, fiber transmission significantly modified the observed polarization correlations in both configurations. The corresponding polarization-correlation curves and reconstructed density matrices before compensation are shown in Figs.~\ref{fig:qhq_results}(a,c) and \ref{fig:fpc_results}(a,c) for the QHQ and FPC measurements, respectively. In both cases, the correlation patterns differ markedly from those obtained for the reference EPS, and the reconstructed density matrices show a redistribution of the state components relative to the target $\left|\Phi^{+}\right\rangle$ state.

This behavior is also reflected in the quantitative results summarized in Table~\ref{tab:combined_results}. Before compensation, the measured visibilities were considerably reduced with respect to those obtained for the reference EPS, demonstrating that the polarization transformations introduced by the fibers altered the correlations observed in the fixed measurement bases. Similarly, the measured CHSH parameters were well below the value obtained for the reference source.

The reduction in the measured visibility and the corresponding change in the CHSH parameter indicate that fiber transmission modified the polarization correlation structure relative to the measurement settings associated with the target $\left|\Phi^{+}\right\rangle$ state. Since the signal and idler photons propagate through separate fiber paths, each photon can undergo an independent polarization transformation. Consequently, when the resulting two-photon state is characterized in a fixed polarization basis, its observed correlations can differ markedly from those expected for the target state, even when a significant degree of entanglement is preserved.

The QST results provide further insight into the nature of these changes. Before compensation, the reconstructed states exhibited low fidelities with respect to the target $\left|\Phi^{+}\right\rangle$ state, whereas the corresponding concurrence and purity remained relatively high, as summarized in Table~\ref{tab:combined_results}. Thus, a low fidelity with respect to the specific target state $\left|\Phi^{+}\right\rangle$ does not necessarily imply a comparable reduction in the degree of entanglement. Instead, the combination of low fidelity with relatively high concurrence and purity is consistent with the distributed state having undergone local polarization transformations that change its representation relative to the fixed $\left|\Phi^{+}\right\rangle$ basis used for characterization.

This distinction is important because, in the ideal case, local unitary transformations can change the representation of a two-photon state and substantially reduce its fidelity with a particular target Bell state without changing its entanglement. Therefore, fidelity with respect to $\left|\Phi^{+}\right\rangle$ alone is not sufficient to distinguish a mismatch between the distributed polarization basis and the chosen target basis from an actual degradation of the underlying entanglement.

Following the path-by-path optimization, the polarization correlations were effectively restored for both compensation configurations, as shown in Figs.~\ref{fig:qhq_results}(b,d) and~\ref{fig:fpc_results}(b,d). The post-compensation correlation curves show the expected polarization dependence, while the reconstructed density matrices exhibit a state structure close to the target $\left|\Phi^{+}\right\rangle$ state. The post-compensation visibilities exceeded $93\%$ in all three measurement bases. The measured CHSH parameters increased to $2.661\pm0.018$ and $2.664\pm0.017$ for the QHQ and FPC configurations, respectively. Both values exceed the classical bound and approach the value obtained for the reference EPS.

The QST results showed a similar restoration of the target state. The reconstructed states exhibited fidelities above $94.5\%$ with respect to $\left|\Phi^{+}\right\rangle$, together with high concurrence and purity. Taken together, the polarization-correlation, Bell--CHSH, and QST measurements demonstrate that the compensation procedures restored the distributed states close to the target $\left|\Phi^{+}\right\rangle$ state.

The post-compensation results obtained with the QHQ and FPC configurations were closely comparable. The recovered visibilities exceeded $93\%$ for both techniques, while the measured CHSH parameters and state fidelities showed only small differences under the present experimental conditions. No clear performance advantage of one technique over the other is evident from the measured polarization visibility, Bell--CHSH, or state-tomography results. The two approaches therefore provide comparable restoration of the target polarization-entangled state while differing primarily in their physical implementation: the QHQ configuration performs polarization control in free space through adjustable waveplate orientations, whereas the FPC provides polarization control directly within the fiber.

An important feature of the present approach is that the polarization transformation introduced along each fiber path is compensated independently before evaluating the final two-photon state. This path-by-path procedure provides a systematic way to correct the polarization change in each distribution channel without relying solely on the final two-photon coincidence signal during the initial optimization. Since the signal and idler photons experience independent polarization transformations during fiber transmission, optimizing each path separately allows the corresponding transformations to be corrected before the final characterization of the distributed state.

The results obtained from the polarization-correlation measurements, Bell--CHSH parameter, and reconstructed quantum states confirm the effectiveness of this procedure. The comparable restoration achieved with the QHQ and FPC implementations further shows that the same path-by-path compensation principle can be realized using either free-space waveplate-based or fiber-based polarization control. This provides practical flexibility in selecting a compensation method based on the requirements of the experimental implementation, including optical configuration and system integration, for laboratory-scale and short-reach fiber-based quantum communication experiments.

\section{Conclusion}

In this work, passive polarization compensation of fiber-distributed polarization-entangled photons was experimentally investigated using QHQ waveplates and a three-paddle fiber polarization controller (FPC). Fiber transmission introduced polarization transformations that significantly altered the measured polarization correlations and changed the reconstructed two-photon state relative to the target $\left|\Phi^{+}\right\rangle$ state.

Both compensation approaches were evaluated under identical experimental conditions, with the two fiber paths optimized independently. After compensation, high polarization correlations and Bell--CHSH violation were recovered, while QST confirmed the restoration of states close to the target $\left|\Phi^{+}\right\rangle$ state, with fidelities approaching that of the reference source. The comparable results obtained with the two methods demonstrate similar recovery of the target polarization-entangled state under the present experimental conditions.

The close agreement between the two methods shows that both passive
approaches can effectively restore the desired polarization correlations
after fiber transmission under the present experimental conditions. During
the several-hour measurement period required for the polarization-correlation,
Bell--CHSH, and quantum-state-tomography measurements, no additional
polarization adjustment was required under laboratory temperature regulation
within approximately $\pm0.5^{\circ}\mathrm{C}$. This observation indicates
that both compensation configurations were sufficiently stable for the
duration of the present measurements. The results also demonstrate the
usefulness of a systematic path-by-path compensation procedure, in which each
optical path is optimized independently before the final two-photon
characterization.

Since the measurements were performed using fixed $2$~m fiber paths under laboratory conditions, longer or dynamically varying links may require periodic or active compensation. The choice between QHQ and FPC can therefore be made based on the requirements of the experimental implementation, such as optical configuration, ease of adjustment, and system integration, rather than a significant difference in the recovered polarization-entanglement characteristics.

\begin{backmatter}

\bmsection{Funding}
The authors acknowledge the support from DIAT(DU) under the grant-in-aid program.

\bmsection{Acknowledgment}
The authors acknowledge useful discussions with P. Kanaka Raju (School of Quantum Technology, DIAT, Pune, India).

\bmsection{Disclosures}
The authors declare no conflicts of interest.

\bmsection{Data availability} All data supporting the findings of this study are included in the article.
\end{backmatter}

\bibliography{sample}

\end{document}